\documentclass{JINST}
\pdfoutput=1 

\usepackage[utf8x]{inputenc}
                     
\usepackage{graphicx}
\usepackage{amsmath}
\usepackage{amssymb}
\usepackage{wasysym}
\usepackage{multirow}
\usepackage{textcomp}
\usepackage{multirow}
\usepackage{subfigure}
\usepackage{footnote}

\title{\boldmath The GeMSE Facility for Low-Background\\ $\gamma$-Ray Spectrometry}

\author{M. v. Sivers$^a$\thanks{Corresponding author.}, B. A. Hofmann$^{b,c}$, \r{A}. V. Ros\'{e}n$^b$, and M. Schumann$^{a,d}$ \\
\llap{$^a$}Albert Einstein Center for Fundamental Physics, University of Bern,\\CH-3012 Bern, Switzerland \\
\llap{$^b$}Institute of Geological Sciences, University of Bern,\\CH-3012 Bern, Switzerland \\
\llap{$^c$}Natural History Museum Bern,\\CH-3005 Bern, Switzerland \\
\llap{$^d$}current institution: Institute of Physics, University of Freiburg,\\D-79104 Freiburg, Germany \\

 E-mail: \email{moritz.vonsivers@lhep.unibe.ch, marc.schumann@lhep.unibe.ch}
}

\abstract{We describe a new high-purity germanium (HPGe) detector setup for low-background $\gamma$-ray spectrometry. The GeMSE facility (Germanium Material and meteorite Screening Experiment) is dedicated to material screening for rare event searches in astroparticle physics as well as to the characterization of meteorites.
It is installed in a medium depth ($\sim$620\,m.w.e.) underground laboratory in Switzerland in a multi-layer shielding and is equipped with an active muon veto.
We have reached a very competitive integral background rate of $(246\pm2)$\,counts/day (100-2700\,keV) and measured a sensitivity of \mbox{$\sim$0.5-0.6\,mBq/kg} for long-lived isotopes from the $^{238}$U/$^{232}$Th chains in a $\sim$1\,kg sample screened for $\sim$27\,days. An extrapolation to higher sample masses and measurement times suggests a maximum sensitivity in the $\mathcal{O}(50)$\,\textmu Bq/kg range. We describe the data analysis based on Bayesian statistics, background simulations, the efficiency calibration and first sample measurements.}


\begin{document}



\section{Introduction}
Low-background $\gamma$-ray spectrometry is a widely used tool to select the very radiopure materials needed for rare event searches in astroparticle physics, e.g., the search for dark matter or neutrinoless double beta decay \cite{leonard08,aprile11}. In a different field, namely meteorite research, $\gamma$-ray spectrometry can be used to determine the terrestrial age of meteorite samples by detecting cosmogenic isotopes \cite{bhandari08,buhl14,evans92}. GeMSE (Germanium Material and meteorite Screening Experiment) \cite{sivers15} is an interdisciplinary project addressing both of these topics. Its goal is to build and operate a highly sensitive screening setup that will be used for the identification of radiopure materials, for the future dark matter experiments XENONnT \cite{aprile14,aprile16} and DARWIN \cite{darwin01,schumann15}, as well as for the identification of recent falls from the Oman meteorite collection \cite{hofmann04,al-kathiri05} hosted at the Natural History Museum Bern. This paper describes the GeMSE facility and its performance.

\section{Detector Description}
\begin{figure}[t!]
\centering
\includegraphics[width=0.6\textwidth]{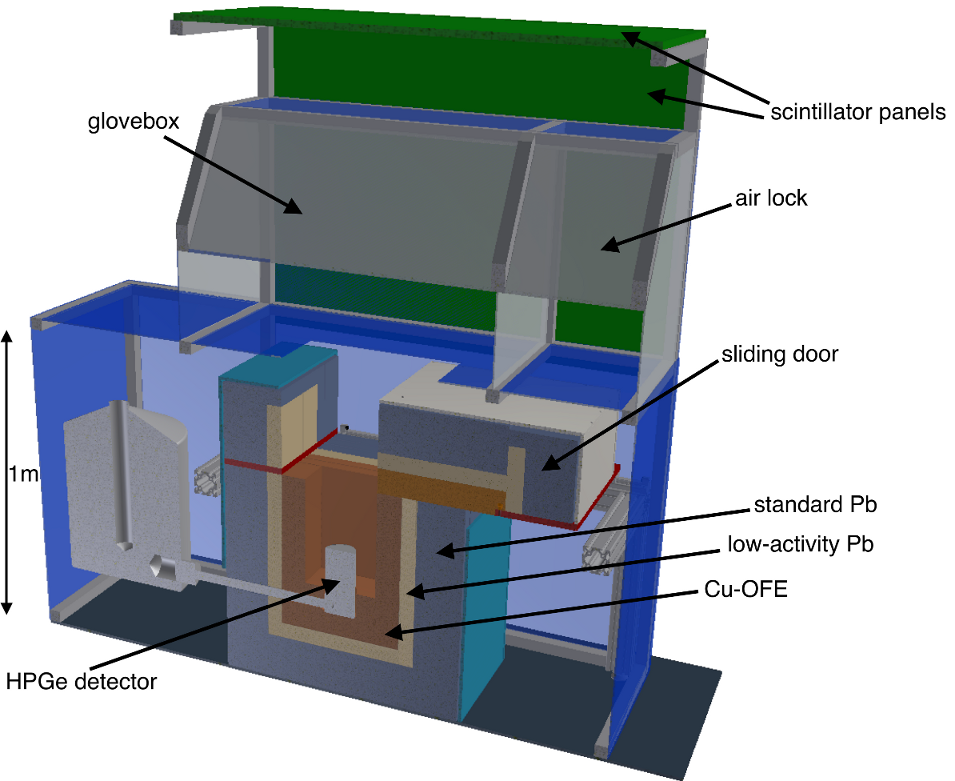}
\caption{Section view of the GeMSE setup. The HPGe detector is surrounded by several layers of shielding and enclosed in a N$_2$-purged glovebox. Plastic scintillator panels on top and on the back serve as muon veto.}
\label{fig:setup}
\end{figure}

GeMSE uses a standard electrode, coaxial, p-type HPGe detector from Canberra \cite{canberra}. The certified relative detection efficiency is 107.7\% and the resolution (FWHM) at 1.33\,MeV ($^{60}$Co) is 1.96\,keV at 4$\mu$s shaping time constant. An ultra-low background U-style cryostat made from oxygen free copper houses the Ge crystal ($\diameter=85$\,mm, $h=65$\,mm, $m\approx2.0$\,kg). Figure~\ref{fig:setup} shows a schematic view of the shielding design which was optimized based on background simulations carried out with Geant4 (v9.6p03) \cite{agostinelli02}. The detector is housed in a large-volume sample cavity of $24\times24\times35$\,cm$^3$. From inside to outside the shielding consists of 8\,cm of oxygen free copper (Cu-OFE, $>$99.99\% purity), 5\,cm of low-activity Pb ($7.2\pm0.5$\,Bq/kg from $^{210}$Pb) and 15\,cm of standard Pb ($91\pm14$\,Bq/kg from $^{210}$Pb). The lead bricks were arranged avoiding any direct line of sight to the detector. The entire shielding is enclosed by a glovebox which is continuously purged with boil-off N$_2$ gas which is guided directly into the sample cavity. This reduces the radon activity from $\sim$40\,Bq/m$^3$ in the laboratory air to a negligible level. The glovebox is equipped with a 2-stage lock system to bring in samples without introducing radon. A sliding door mechanism allows the shielding to be opened by hand inside the glovebox. The setup is located in the Vue-des-Alpes underground laboratory near Neuch\^{a}tel (Switzerland) \cite{gonin03}. It provides a rock overburden of 235\,m ($\sim$620\,m.w.e.), reducing the muon flux by a factor of $\sim$2000. To further lower the background from cosmic-ray muons, two plastic scintillator panels are used as muon veto. The panels, each with an area of $105\times140$\,cm$^2$, are placed above and behind the glovebox.

The HPGe detector signal is read-out by a 14-bit digital MCA (CAEN DT5781A) with real-time digital pulse processing \cite{caen}, storing the pulse height and time stamp of every event. Therefore, unstable periods where, e.g., the muon veto or the N$_2$-purge were not fully operational, can be removed from the data. Detector signals within a 10\,\textmu s window after a muon-veto trigger are discarded, introducing a negligible dead time of $\sim$0.5\%. 

The GeMSE installation is equipped with a slow control system \cite{slowcontrol} which monitors and controls various parameters of the detector and its environment. These include the N$_2$ purge flow into the glovebox, the radon activity in the laboratory, the trigger rate of the muon veto as well as bias voltage and leakage current of the HPGe detector. Together with an automatic LN$_2$ refilling system this allows a fully remote operation of the detector for $\sim$3 weeks.

To minimize cosmogenic activation the HPGe detector and the copper for shielding were stored in an underground laboratory ($\sim$70\,m.w.e.) at the University of Bern \cite{tieflabor} before installation. All shielding materials and the copper cryostat of the HPGe detector were carefully cleaned to remove residual surface contaminations. The cleaning of the detector cryostat was carried out under cleanroom (ISO 6) conditions. All materials were degreased with acetone and rinsed with deionized water. The low-activity lead and all copper was additionally etched using different acid solutions. The copper etching procedure was adapted from \cite{zuzel12} and is known to remove $^{210}$Pb from $^{222}$Rn plate-out. During assembly of the setup in the underground laboratory any new contamination, e.g. from dust, was minimized. 

\section{Detection Efficiency}

\begin{figure}[b!]
\centering
\includegraphics[width=0.8\textwidth]{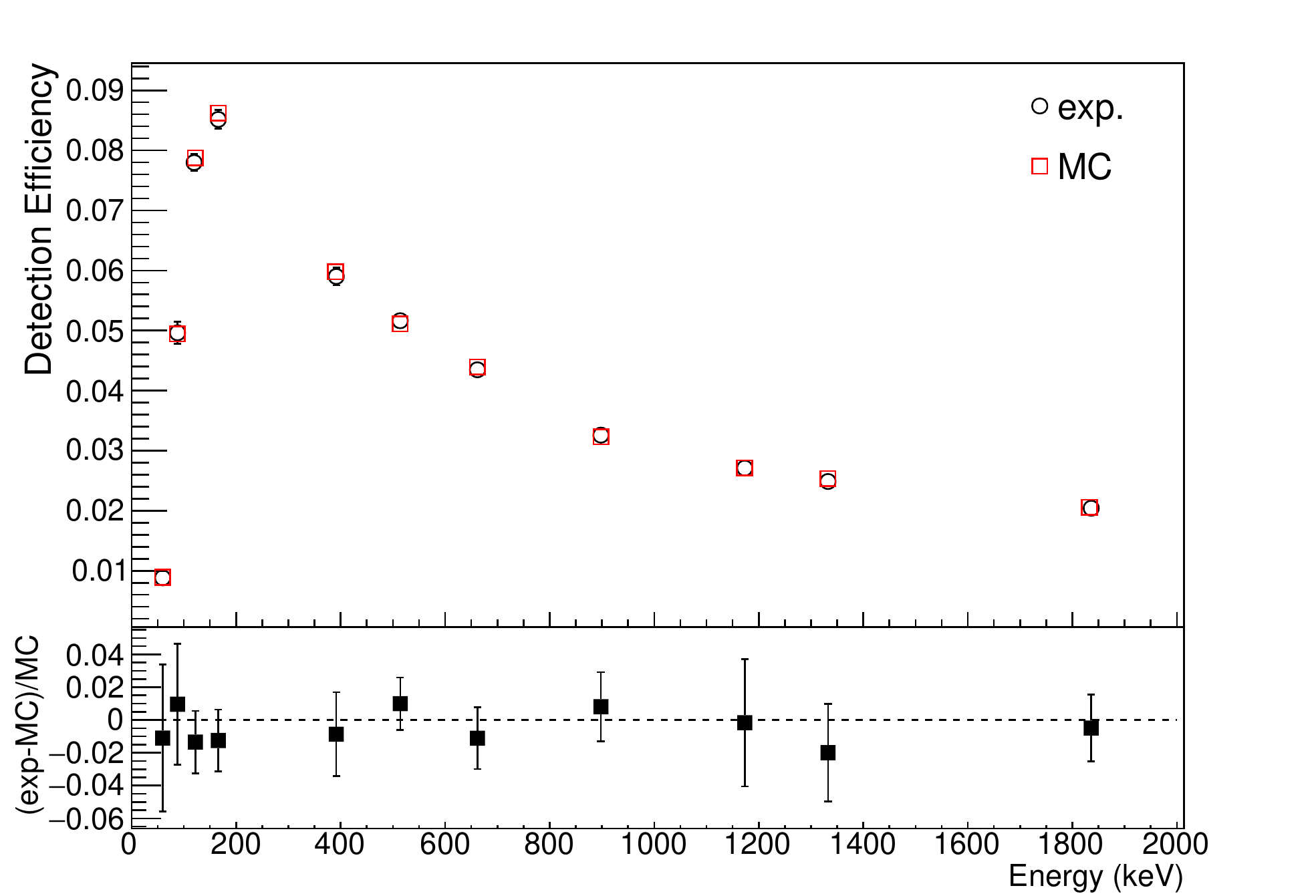}
\caption{Comparison between measured (exp.) and simulated (MC) detection efficiency for several $\gamma$-lines from a disc-shaped source ($\diameter=74$\,mm, $h=28$\,mm) with certified activities.}
\label{fig:efficiency}
\end{figure}

Measurements to determine the detection efficiency of the HPGe detector, which is related to the thickness of the dead layer from the Li-diffused n$^+$-contact, were performed before and after its underground installation using a disc-shaped source ($\diameter=74$\,mm, $h=28$\,mm) containing several radionuclides dispersed inside a silicone resin (CBSS2 \cite{cbss2}). The activity of each radionuclide was certified by the Czech Metrology Institute with an uncertainty between $\sim$1-2\% depending on the nuclide. The detection efficiency $\epsilon$ for each $\gamma$-line was calculated as
\begin{equation}
\epsilon=\frac{R}{A\cdot f},
\end{equation}
where $R$ is the observed rate in the photopeak, $A$ is the activity of the nuclide and $f$ the $\gamma$-emission probability.
The measured efficiencies were compared to those obtained from a Geant4 simulation of the setup, performed with dead layers of different thickness. The best agreement between measurement and simulation was found with a dead layer thickness of $(0.67\pm0.01)$\,mm. This value is in excellent agreement with a dead layer thickness of $(0.65\pm0.05)$\,mm that has been measured using the peak ratio of a $^{133}$Ba source \cite{sivers15}. Figure~\ref{fig:efficiency} shows a comparison between the simulated and measured detection efficiency using the CBSS2 source. The deviations are smaller than 2\% and agree within the 1$\sigma$ errors. The dead layer and the entrance window of the cryostat limit the useful range of the detector to energies $>$50\,keV. This is sufficient for all cosmogenic and primordial nuclides used in a standard sample analysis, both for meteorites and low-background materials.

\section{Background Measurements}

\begin{figure}[b!]
\centering
\includegraphics[width=0.8\textwidth]{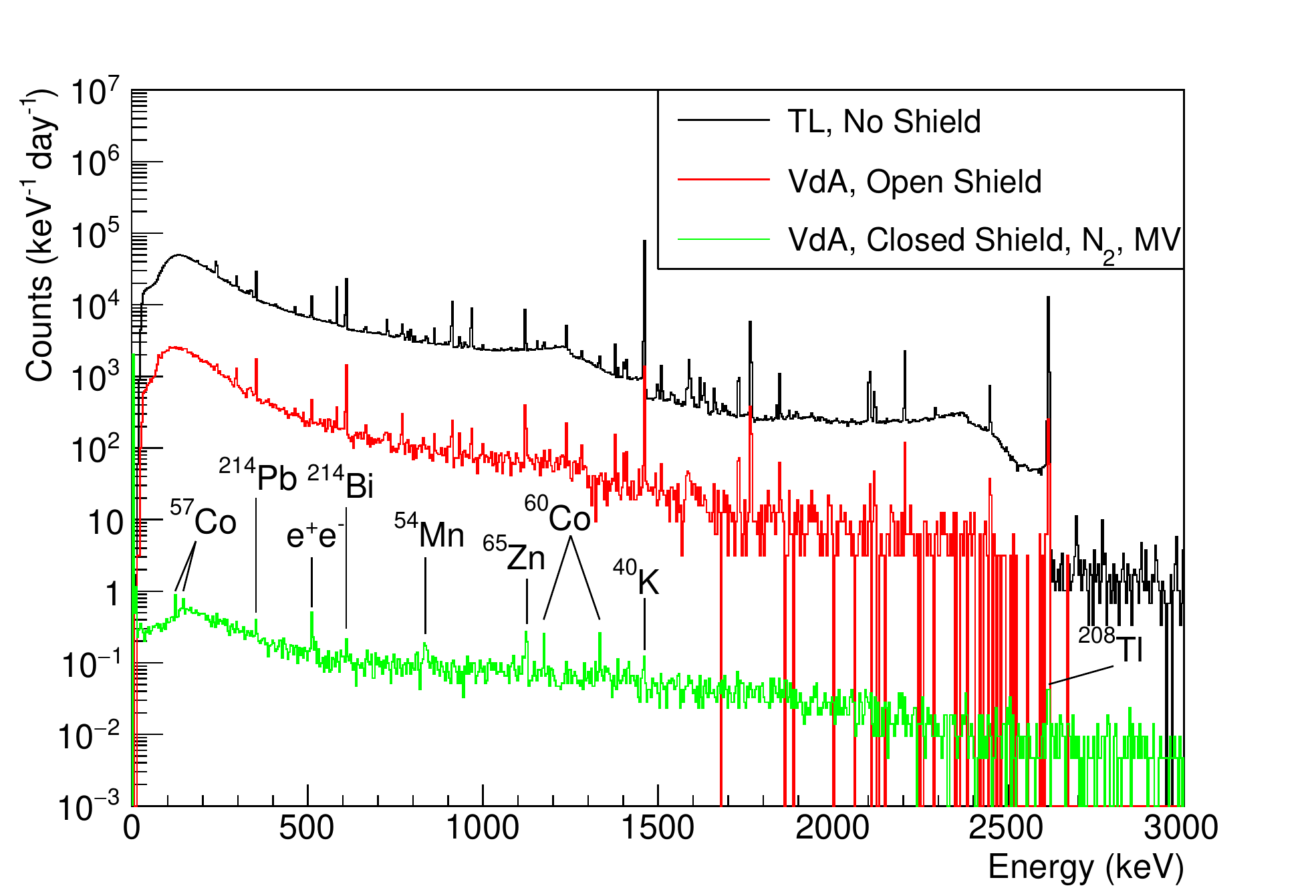}
\caption{GeMSE background spectra taken at different locations in different shielding configurations. The uppermost curve was recorded without any shielding in a shallow underground laboratory ("Tiefenlabor", TL) at the University of Bern. The middle curve was taken at the Vue-des-Alpes (VdA) underground laboratory with open shield door. The lowermost curve corresponds to the final configuration with closed shielding, N$_2$ purging and active muon veto. The total background reduction is almost 5 orders of magnitude.}
\label{fig:background_comparison}
\end{figure}

Figure \ref{fig:background_comparison} shows the GeMSE background spectrum recorded at two different locations and with different shielding configurations. The first spectrum was taken in the shallow underground laboratory ($\sim$70\,m.w.e.) at the University of Bern without any shielding. The second one was recorded after installation at the Vue-des-Alpes underground laboratory ($\sim$620\,m.w.e.), however, with the shielding's sliding door open (see figure~\ref{fig:setup}). The last spectrum was acquired over $\sim$54\,days in the same configuration but with closed shielding, N$_2$ purging of the glovebox at 2-10\,slpm and with the muon veto in operation.
In this configuration the integral background rate (100-2700\,keV) is $(246\pm2)$\,counts/day. This is comparable to the most sensitive screening facilities in the world \cite{laubenstein04} and matches our background goal of 250\,counts/day which was based on initial background simulations \cite{sivers15}. The moun veto reduces the integral background rate (100-2700\,keV) by $\sim$20\%. This small reduction factor is caused by the low geometrical coverage which, however, guarantees easy access to the glovebox. The background rates from cosmogenic and primordial radionuclides are summarized in table~\ref{tab:bkg_counts}. They were obtained with the Feldman \& Cousins procedure \cite{feldman97}, using the count rate in the $\pm3\sigma$ region around the peak and the averaged background from the regions to the left and right of the peak. For all isotopes except the cosmogenic $^{57}$Co and $^{65}$Zn, the count rates are below 1\,count/day. From $^{57}$Co two peaks are visible: a peak at 122\,keV which originates from external contamination and a peak at 144\,keV from an internal contamination of the Ge crystal. For comparison, the table also shows the background rates of the Gator \cite{gator} and GeMPI \cite{heusser06} HPGe spectrometers located at the Gran Sasso underground laboratory (LNGS, Italy) at a depth of 3600\,m.w.e.

\begin{table}[t!]
\centering
\caption{\label{tab:bkg_counts}Background peak count rates of GeMSE for common radioisotopes. The last row gives the integral count rate in the energy region 100-2700\,keV. For comparison, we also show the count rates of two other low-background facilities installed at LNGS. Limits are given at 90\% confidence level (C.L.), uncertainties are at 1$\sigma$.}
\smallskip
\begin{tabular}{|c|c|ccc|}
\hline 
Energy (keV) & Chain/Isotope &  \multicolumn{3}{c|}{Count Rate (day$^{-1}$)} \\
& & GeMSE & Gator \cite{gator} & GeMPI \cite{heusser06} \\
\hline 
122 & $^{57}$Co (ext.) & $1.6\pm0.2$ & - & - \\
144 & $^{57}$Co (int.) & $1.1\pm0.2$ & - & - \\
1125 & $^{65}$Zn & $1.2\pm0.2$ & - & - \\
1173 & $^{60}$Co & $0.84\pm0.15$ & $0.5\pm0.1$ & $0.26\pm0.06$ \\
1333 & $^{60}$Co & $0.84\pm0.15$ & $0.5\pm0.1$ & $0.21\pm0.05$ \\
662 & $^{137}$Cs & $<0.03$ & $0.3\pm0.1$ & $0.34\pm0.16$ \\
1461 & $^{40}$K & $0.23\pm0.10$ & $0.5\pm0.1$ & $0.52\pm0.07$ \\
239 & $^{232}$Th/$^{212}$Pb & $0.34\pm0.17$ & $<0.5$ & - \\
583 & $^{232}$Th/$^{208}$Tl & $0.17\pm0.10$ & - & $\leq0.13$ \\
911 & $^{232}$Th/$^{228}$Ac & $<0.14$ & $<0.5$ & - \\
2615 & $^{232}$Th/$^{208}$Tl & $0.27\pm0.08$ & $0.2\pm0.1$ & $0.11\pm0.03$ \\
352 & $^{238}$U/$^{214}$Pb & $0.67\pm0.17$ & $0.7\pm0.3$ & $\leq0.14$ \\
609 & $^{238}$U/$^{214}$Bi & $0.51\pm0.14$ & $0.6\pm0.2$ & $\leq0.15$ \\
1120 & $^{238}$U/$^{214}$Bi & $<0.02$ & $0.3\pm0.1$ & - \\
1765 & $^{238}$U/$^{214}$Bi & $0.14\pm0.08$ & $0.08\pm0.06$ & - \\
\hline
100-2700 & integral & $246\pm2$ & $226\pm1$ &  $41\pm1$\footnotemark \\
\hline
\end{tabular}
\end{table}
\footnotetext{The integral count rate in \cite{heusser06} is given for the energy range 100-2730\,keV.}

\section{Analysis Method}
\label{sec:analysis}
The analysis to determine the activity of a sample is carried out within a Bayesian framework based on the Bayesian Analysis Toolkit (BAT) \cite{caldwell08}. Such sophisticated analysis is especially important for samples with very low activities, due to the small counting statistics. To obtain the activity for a specific isotope we simultaneous fit all prominent $\gamma$-lines from this isotope, each in a $\pm5\sigma$ region around the peak, in the background and sample spectrum. The fit models consist of template histograms and the free parameters in the fit are the amplitudes of these templates. To evaluate the significance of a signal the fit is performed with a "background-only" and a "background+signal" model.
In the "background-only" fit all energy bins $x$ of the background spectrum around the $i$th $\gamma$-line are fitted with the model
\begin{equation}
B_{i}^{\text{bkg}}(x)=a_i^\text{bkg}G_i^\text{bkg}(x)+c_i^\text{bkg} \,,
\end{equation}
where $G_i^\text{bkg}(x)$ is a normalized Gaussian, $a_i^\text{bkg}$ is the number of counts in the Gaussian peak and $c_i^\text{bkg}$ is a constant term modeling the background under the peak.
For the sample spectrum the model reads
\begin{equation}
B_{i}^{\text{samp}}(x)=a_i^\text{bkg}G_i^\text{bkg}(x)+c_i^\text{samp} \,.
\end{equation}
Note that the number of counts $a_i^\text{bkg}$ in the Gaussian peak is a common parameter for both models but they have different constants $c_i$. This accounts for possible different Compton backgrounds in the two spectra, even if there is no signal in the fit region.
In the "signal+background" case the background spectrum is fitted with the same model as the "background-only" case
\begin{equation}
S_{i}^{\text{bkg}}(x)=a_i^\text{bkg}G_i^\text{bkg}(x)+c_i^\text{bkg} \,,
\end{equation}
while the fit to the sample spectrum contains an additional Gaussian term for the signal peak
\begin{equation}
S_{i}^{\text{samp}}(x)=a_i^\text{bkg}G_i^\text{bkg}(x)+a^\text{samp}\epsilon_i(1+\delta_i \Delta \epsilon)G_i^\text{samp}(x)+c_i^\text{samp} \,,
\end{equation}
where $a_\text{samp}$ is the number of decays in the sample. The factor $\epsilon_i$ is the product of the detection efficiency and emission probability of the $i$th $\gamma$-line. The detection efficiency is determined by a Geant4 simulation of the GeMSE setup including the sample. The term $(1+\delta_i \Delta \epsilon)$ accounts for the systematic uncertainty of the detection efficiency. For samples with complicated geometry or not well known density we usually conduct efficiency simulations with varying shapes or densities. The difference between the different simulated detection efficiencies is taken as systematic uncertainty. Usually, the relative uncertainty ranges between $\Delta \epsilon \approx 2-10\%$. The nuisance parameter $\delta_i$ is a free parameter in the fit with a normal distribution as prior.
The free parameters of the "background-only" and the "background+signal" fits for $N$ considered $\gamma$-lines are
\begin{subequations}
\begin{align}
\vec{\lambda}_\text{B} & =\{c_0^\text{bkg},c_0^\text{samp},a_0^\text{bkg},...,c_N^\text{bkg},c_N^\text{samp},a_N^\text{bkg}\} \,,
\\
\vec{\lambda}_\text{S} & =\{c_0^\text{bkg},c_0^\text{samp},a_0^\text{bkg},\delta_0,...,c_N^\text{bkg},c_N^\text{samp},a_N^\text{bkg},\delta_N,a_\text{samp}\} \,.
\end{align}
\end{subequations}
All parameters are constrained to physically allowed positive values.
The total posterior probability distribution is given by
\begin{subequations}
\begin{align}
P(\vec{\lambda} \mid \vec{D}) & =  \frac{P(\vec{D}\mid \vec{\lambda}) P_{0}(\vec{\lambda})}{\int P(\vec{D}\mid \vec{\lambda}) P_0(\vec{\lambda})\,\text{d}\vec{\lambda}} \,,
\end{align}
with
\begin{equation}
P(\vec{D}\mid \vec{\lambda}) = \prod\limits_{i=1}^{N}P(\vec D_i^\text{samp} \mid \vec{\lambda}) P(\vec D_i^\text{bkg} \mid \vec{\lambda}) \,.
\end{equation}
\end{subequations}
$\vec D_i^\text{samp}$ and $\vec D_i^\text{bkg}$ are the $\pm5\sigma$ regions around the $i$th $\gamma$-line of the sample and background spectrum, respectively. In the calculation of the likelihood $P( \vec{D} \mid \vec{\lambda})$, Poissonian statistical uncertainties are assumed for each bin entry of the data histograms $\vec D$. The priors $P_0(\vec{\lambda})$ are usually assumed to be flat. If, however, the activity of an isotope is known, e.g., from a previous measurement, one may choose the prior for the parameter $a_\text{samp}$ accordingly. 

The Bayes Factor ($BF$) is used to decide whether or not there is a signal in the spectrum. It is defined as the ratio of the posterior probabilities of the "background-only" $B$ and the "signal+background" fit $S$ under the assumption of equal priors
\begin{equation}
BF=\frac{P(B \mid \vec{D})}{P(S \mid \vec{D})} \,.
\end{equation}
The choice of a threshold on $BF$ below which a signal detection is claimed depends on the type of measurement. In material screening measurements it is more conservative to get a false detection than a false negative. Therefore, we choose a rather high threshold of $BF\leq0.33$ for a signal detection (defined in \cite{kass95} as "positive evidence"). In case of a meteorite screening we require $BF\leq0.05$ (defined in \cite{kass95} as "strong evidence") to claim the detection of a cosmogenic isotope. 

In case of a signal detection we calculate the activity and its uncertainty from the mode and central 68\% interval of the marginalized posterior probability distribution 
\begin{equation}
P(a_\text{samp} \mid \vec{D}) = \int P(\vec{\lambda} \mid \vec{D}) \text{d}\vec{\lambda}\vert_{\lambda_i\neq a_\text{samp}} \,.
\end{equation}
When there is no detectable signal, an upper limit on the activity is calculated. This is done by integrating the marginalized posterior probability distribution of the signal activity. For example, the 95\% credible interval (C.I.) upper limit $a_\text{lim}$ on the number of counts from the sample is given by
\begin{equation}
\frac{\int_0^{a_\text{lim}} P(a_\text{samp} \mid \vec{D}) \text{d}a_\text{samp}}{\int_0^\infty P(a_\text{samp} \mid  \vec{D}) \text{d}a_\text{samp}}=0.95 \,.
\end{equation}

\section{Background Analysis}

\begin{figure}[t!]
\centering
    \subfigure{\includegraphics[width=0.7\textwidth]{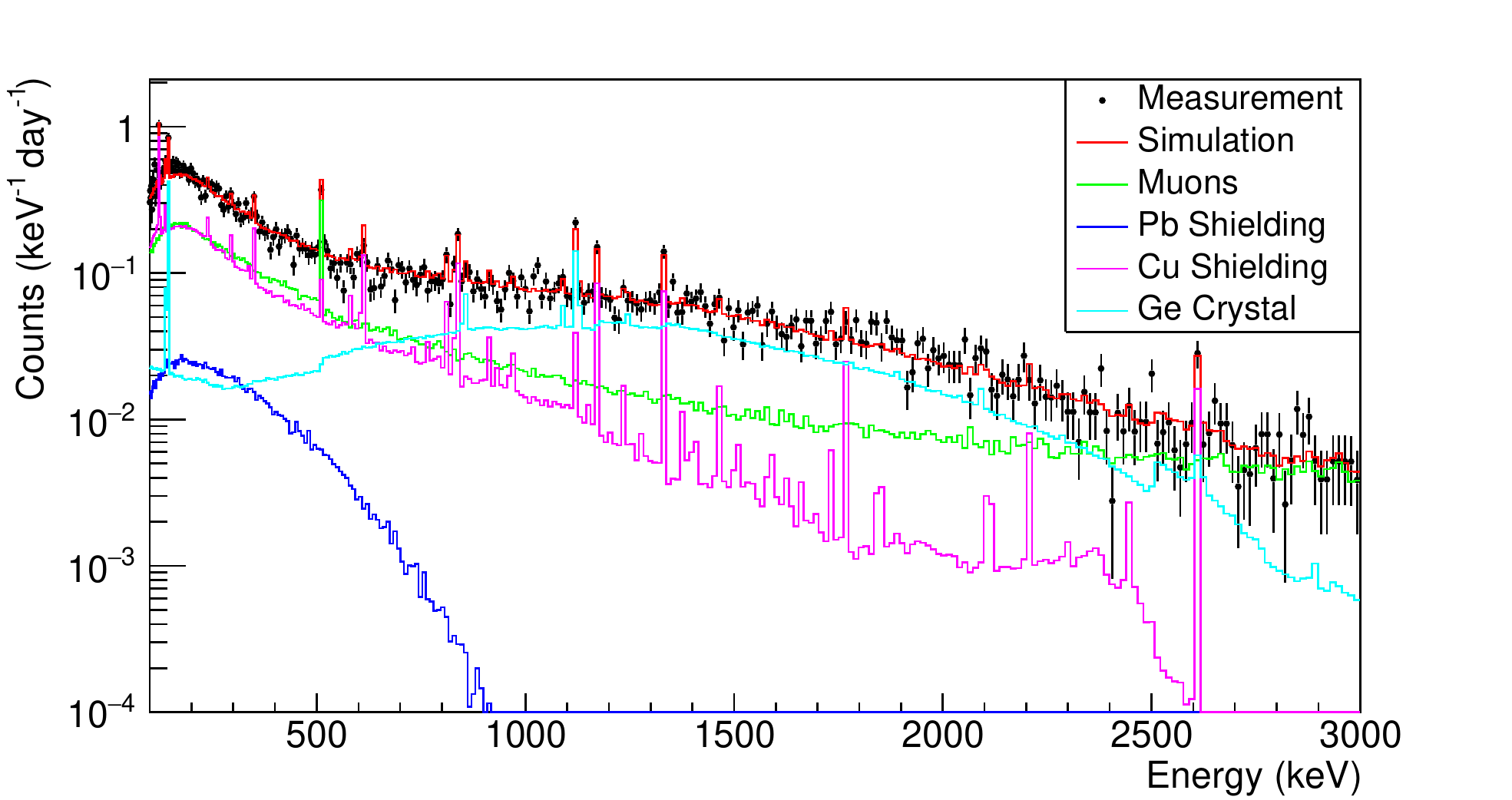}}
    \subfigure{\includegraphics[width=0.7\textwidth]{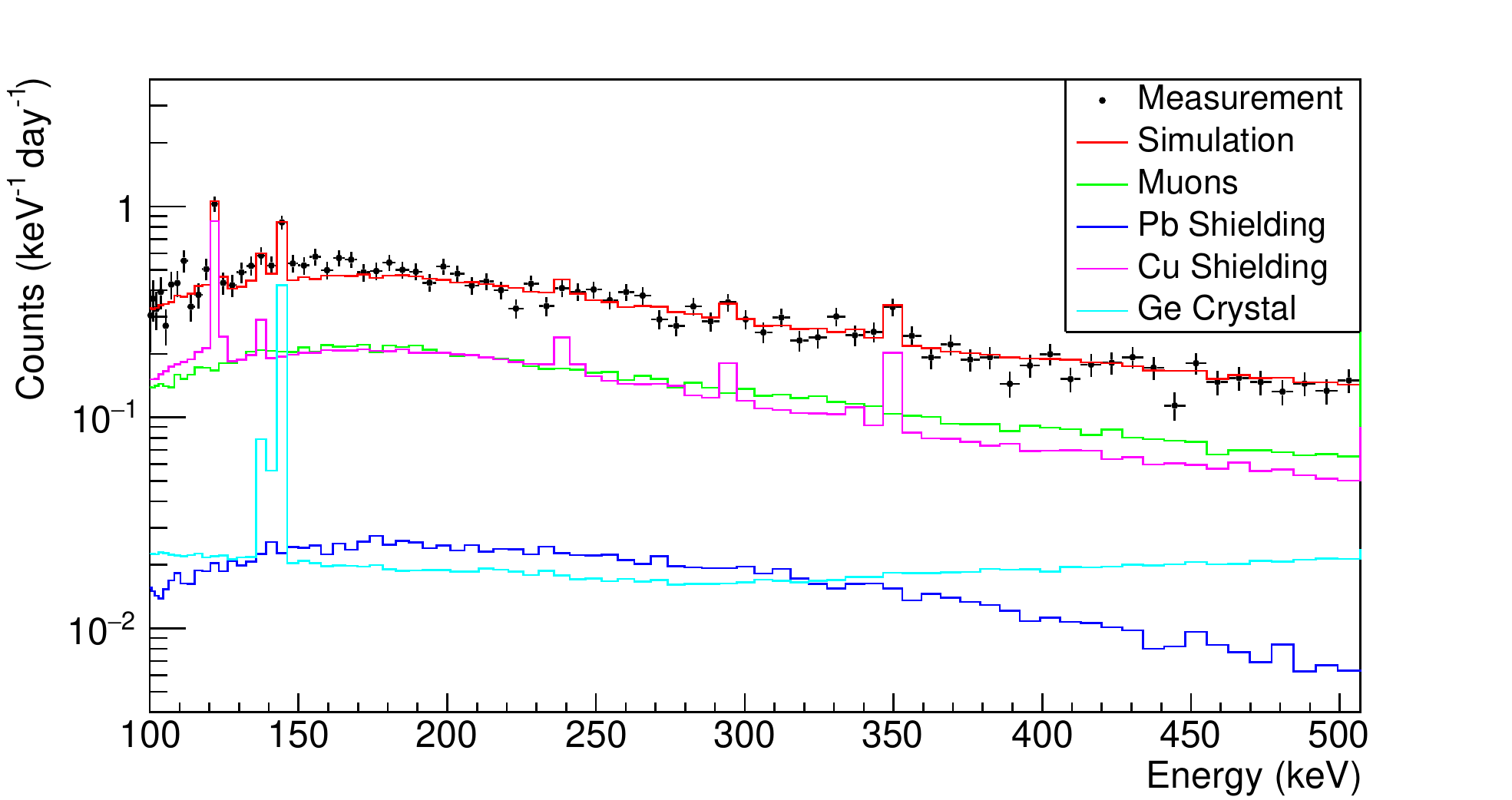}}
    \caption{Best fit to the measured background spectrum. The activities extracted from the fit are shown in table~2. The bottom panel shows a zoom into the low-energy region.}
    \label{fig:bkg_fit}
\end{figure}

\begin{table}[b!]
\centering
\caption{\label{tab:bkg_fit}Radioactive contaminations in the HPGe detector and the materials of the shield, as well as the flux from cosmic ray muons. The values were extracted by fitting simulated spectra to the measured background spectrum. For short-lived isotopes the activity at the end of the measurement is given. Uncertainties and upper limits are given at 68\% C.I. and 95\% C.I., respectively. }
\vspace{0.2cm}
\small
\begin{tabular}{|l|cccccccccc|}
\hline
& & & & & & & & & & \\[-0.4cm]
Source &  \multicolumn{10}{|c|}{Specific Activity of Isotope (\textmu Bq/kg)} \\
& $^{65}$Zn & $^{68}$Ge & $^{60}$Co & $^{58}$Co & $^{57}$Co & $^{56}$Co & $^{54}$Mn & $^{238}$U & $^{232}$Th & $^{40}$K \\[0.1cm]
\hline 
& & & & & & & & & & \\[-0.4cm]
Ge Crystal & 77${+15 \atop -13}$ & 313${+17 \atop -23}$ & $<$118 & $<$3.5 & 11${+2 \atop -1}$ & $<$43 & $<$23 & -- & -- & -- \\[0.1cm]
Cu Shield & -- & -- & 43${+7 \atop -5}$ & $<$26 & 374${+48 \atop -39}$ & $<$7.6 & $<$56 & 104${+13 \atop -15}$ & 42${+9 \atop -10}$ & $<$196 \\[0.1cm] \hline
& & & & & & & & & & \\[-0.4cm]
Pb Shield & \multicolumn{10}{|l|}{$^{210}$Pb activity $<$8.2\,Bq/kg} \\[0.1cm] 
Muons & \multicolumn{10}{|l|}{Flux = $1.19{+0.07 \atop -0.09} \times 10^{-5}$\,cm$^{-2}$\,s$^{-1}$ } \\[0.1cm] \hline
\end{tabular}
\end{table}

In order to determine the origin of the remaining background we used a Geant4 simulation of the GeMSE setup including the detector, shielding and muon veto. Using the ``Shielding'' physics list we simulated the background from cosmic-ray muons, primordial and cosmogenic nuclides in the shielding and detector materials as well as that from residual $^{222}$Rn in the sample cavity. The simulated background spectra were fitted to the measured spectrum using a similar code as for the sample analysis described in section~\ref{sec:analysis}. The previously measured activity of $(7.2\pm0.5)$\,Bq/kg was used as a Gaussian prior for the $^{210}$Pb activity of the inner lead shielding. For the muon background an integral flux of $7.1\times10^{-6}$\,cm$^{-2}$\,s$^{-1}$, calculated from the known rock overburden, was used as Gaussian prior with an uncertainty of 30\%. Uniform priors were used for all other activities. The simulated muon background already takes into account the effect of the muon veto, discarding all hits in the HPGe detector within 10\,\textmu s after an energy deposition $>$1\,MeV in one of the scintillator panels. Contributions from $^{222}$Rn in the sample cavity ($\sim$0.3\,mBq/m$^3$) and contaminations in the detector's Cu cryostat ($\sim$0.8\,mBq/kg) were found to be negligible, therefore, they were not included in the final fit. 

Table \ref{tab:bkg_fit} summarizes the activities extracted from the fit. A comparison between data and simulation is shown in figure~\ref{fig:bkg_fit}. The p-value of the fit, calculated as described in \cite{beaujean11}, is 0.02. Considering the large number of data points, this indicates that the model describes the data well. The dominant background contribution at energies $\lesssim$700\,keV comes from muons and $^{238}$U and $^{232}$Th contaminations in the Cu shielding. In addition, there are lines at 122\,keV and 144\,keV from $^{57}$Co in the Cu shielding and Ge crystal, respectively. In the energy range $\sim$700-2300\,keV, the dominant background originates in cosmogenic activation of the Ge crystal: the $\beta$-spectrum of $^{68}$Ga (short-lived daughter of $^{68}$Ge, Q-value 2921\,keV) is clearly visible in the data. The dominating background above 2300\,keV is due to muons, however the flux extracted from the fit is about 70\% higher than expected. This can be explained by the lack of knowledge of the exact profile and density of the rock overburden and that the efficiency of the muon veto might be overestimated in the simulation. 

We note that there is some degeneracy in the background fit: In general, the spectra from contaminations in the Cu shielding and detector Cu cryostat are very similar. This also holds for $^{238}$U in the Cu shielding and $^{222}$Rn in the sample cavity. It is therefore possible that the rather high $^{57}$Co activity attributed to the Cu shielding partially originates from the detector cryostat. Similarly, some of the $^{238}$U activity might come from residual $^{222}$Rn in the sample cavity.
Since a significant part of the background originates from rather short-lived cosmogenic isotopes, we expect that background rate will decrease to $\sim$190 (160)\,counts/day after 1 (5)\,years.


\section{Sample Measurements}

\begin{table}[b!]
\centering
\caption{\label{tab:meteorite}Activities and concentration of cosmogenic (a) and primordial isotopes (b) in the 50.8\,g Boumdeid (2011) meteorite sample measured with GeMSE (this work) and from a previous measurement by Buhl et al.~\cite{buhl14}. The activity is given in decays per minute (1\,dpm$=16.7$\,mBq). Upper limits from this work and \cite{buhl14} are given at 95\%~C.I. and 68\%~C.L., respectively. Uncertainties on detected activities are at 68\% C.I. (this work) and 68\% C.L. (\cite{buhl14}).}\vspace{0.2cm}
\small
\begin{tabular}{|cc|cc|c|c|cc|}
\multicolumn{5}{l}{(a) Cosmogenic Isotopes} & \multicolumn{2}{l}{(b) Primordial Isotopes} \\
\cline{1-4} \cline{6-8} 
& & & & & & &\\[-0.4cm]
Isotope & $T_{1/2}$ (y) & \multicolumn{2}{|c|}{Specific Activity (dpm/kg)} & & Isotope & \multicolumn{2}{|c|}{Concentration (ng/g)} \\
& & GeMSE & Buhl et al. \cite{buhl14} & & & GeMSE & Buhl et al. \cite{buhl14} \\
\cline{1-4} \cline{6-8}
& & & & & & & \\[-0.4cm]
$^{54}$Mn & 0.854 & 72$+13 \atop -8$ & $71.7 \pm 7.3$ & & $^{238}$U  & 10.7$+0.6 \atop -0.6$  & $14\pm2$ \\[0.1cm]   
$^{22}$Na & 2.603 & 86$+6 \atop  -5$ & $91.9 \pm 9.5$ & & $^{232}$Th & 34$+2 \atop -2$ & $38\pm5$ \\[0.1cm]
 \cline{6-8}
$^{60}$Co & 5.217 & $<$1.02 & $<$0.47 & &  & \multicolumn{2}{|c|}{Total K Concentration (mg/g)} \\[0.1cm]
 \cline{6-8}
 & & & & & & &\\[-0.4cm]
$^{44}$Ti & 60.0 & $<$0.84 & $<$2.5 & & $^\text{nat}$K & 0.80$+0.05 \atop -0.04$ & $0.89\pm0.09$ \\ [0.1cm]
 \cline{6-8}
$^{26}$Al & 7.17$\times$10$^5$ & 51$+3 \atop -2$  & $57.1 \pm 6.1$ &  \multicolumn{4}{c}{} \\[0.1cm]
\cline{1-4}
\end{tabular}
\end{table}

\begin{table}[b!]
\centering
\caption{\label{tab:HVplug}Radioactive contamination in a batch of custom-made low-radioactivity high voltage connectors (20 pieces of 52\,g each). Upper limits are given at 95\%~C.I., uncertainties on detected activities are at 68\% C.I.}\vspace{0.2cm}
\small
\begin{tabular}{|cc|ccccccc|}
\hline
\multicolumn{2}{|c|}{Isotope} & $^{238}$U & $^{226}$Ra & $^{228}$Ra & $^{228}$Th & $^{60}$Co & $^{40}$K & $^{137}$Cs \\ 
\hline 
&&&&&&&& \\[-0.4cm]
\multirow{2}{*}{Specific Activity} & (mBq/kg) &  $<8.5$ & $<0.45$ & $<0.55$ & $<0.57$  & 1.2$+0.2 \atop -0.2$ & 4.0$+1.8 \atop -1.4$ & $<0.37$ \\ [0.1cm] 
 & (\textmu Bq/pc) & $<$440 & $<$24  & $<$29 & $<$30 & 64$+12 \atop -10$  & 206$+96 \atop -73$   & $<$19 \\ [0.1cm] 
\hline
\end{tabular}
\end{table}

\begin{figure}[t!]
\centering
    \subfigure{\includegraphics[width=0.7\textwidth]{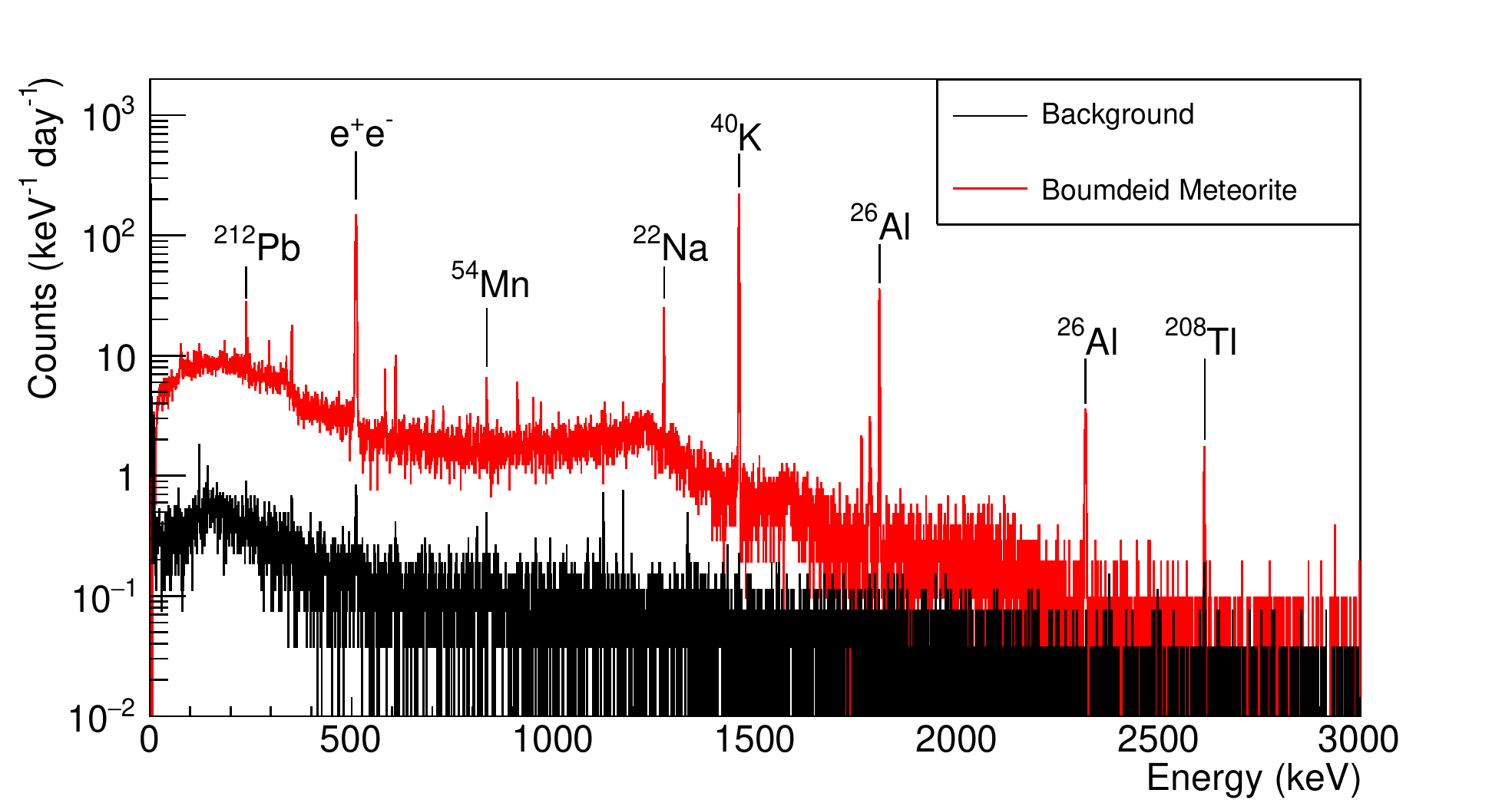}}
    \subfigure{\includegraphics[width=0.7\textwidth]{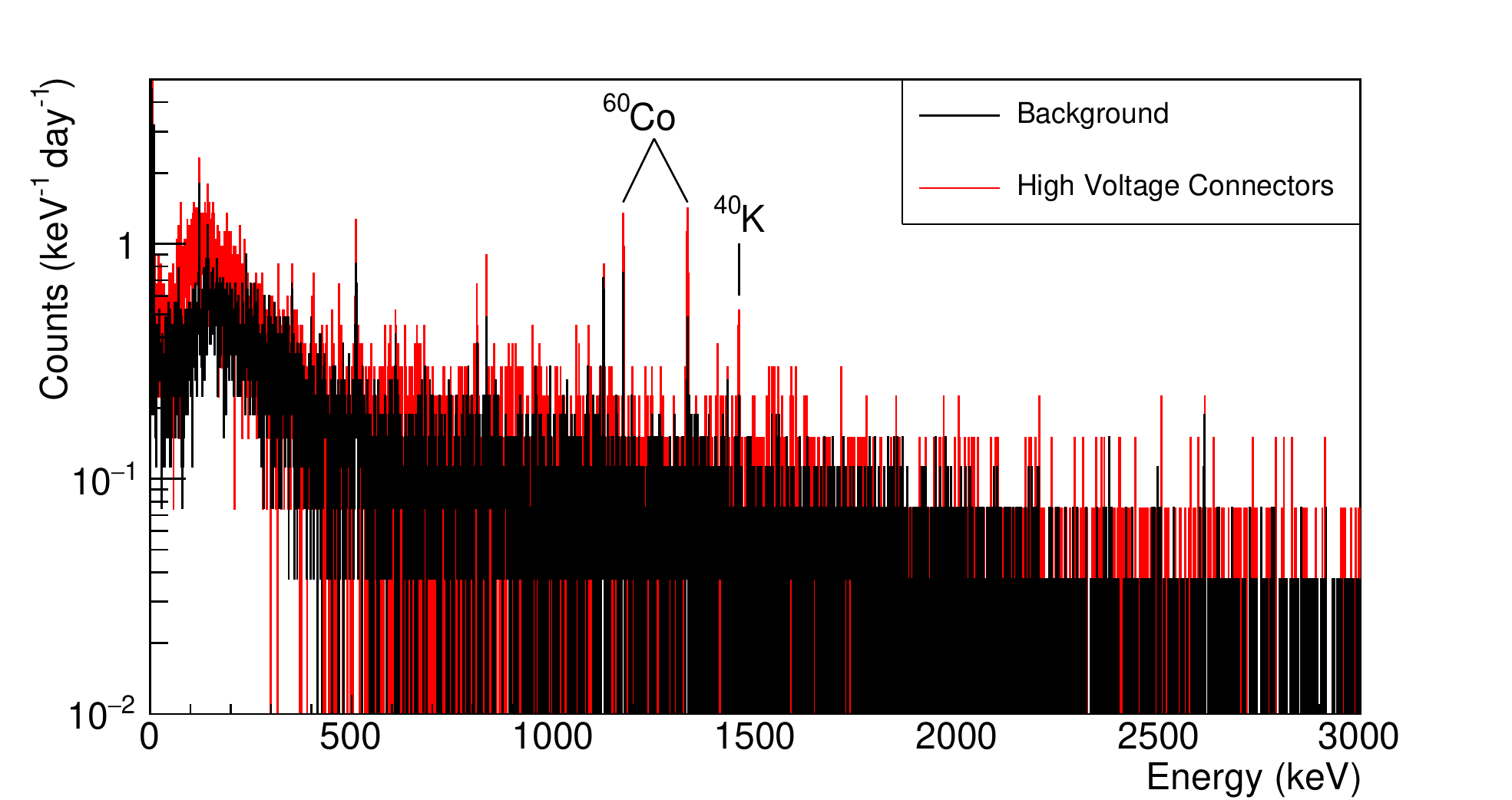}}
    \caption{Sample and background spectra for a meteorite (top) and a batch of custom-made low-background high voltage connectors (bottom).}
    \label{fig:sample_spectra}
\end{figure}

We present the results for two typical samples measured in the GeMSE setup. The first is a fragment of the Boumdeid (2011) meteorite \cite{ruzicka14} which fell on September 14, 2011 near Boumdeid, Mauritania. The sample, an ordinary chondrite of type L6, has a mass of 50.8\,g and was measured for 434\,h. Figure~\ref{fig:sample_spectra} (top) shows the measured spectrum. The detection efficiency was determined by a Geant4 simulation using the average L-chondrite major element composition from \cite{wasson88} and the density from \cite{britt03}. The activities of relevant cosmogenic radionuclides in the sample, calculated back to the time of fall, are listed in table~\ref{tab:meteorite}(a). Table~\ref{tab:meteorite}(b) summarizes the concentrations of primordial isotopes. The same sample was previously measured in the STELLA (SubTerranean Low Level Assay) facility at LNGS \cite{buhl14}. The activities measured with GeMSE are in good agreement with those.

The second sample is a batch of 20 custom-made high voltage connectors, similar to the ones used in the XENON1T experiment, with a total mass of 1.04\,kg. The sample was measured for 648\,h, the spectrum is presented in figure~\ref{fig:sample_spectra} (bottom). The results for the activities of common radioisotopes are shown in table~\ref{tab:HVplug}.

\section{Summary and Conclusion}
We have built the new highly-sensitive GeMSE facility for low-background $\gamma$-ray spectrometry in a medium-depth underground laboratory in Switzerland. The HPGe detector is surrounded by several layers of shielding, enclosed in a N$_2$-purged glovebox and equipped with an additional muon veto. We have reached an integrated background rate (100-2700\,keV) of $246\pm2$\,counts/day, which is comparable to the most sensitive screening facilities in the world \cite{laubenstein04}. Since background simulations have shown that short-lived cosmogenic isotopes contribute significantly to the observed background, we expect the background rate to decrease even further in the next years. For a $\sim$1\,kg low-background sample measured for $\sim$27\,days, we have reached a sensitivity of \mbox{$\sim$0.5-0.6\,mBq/kg} for long-lived isotopes from the $^{238}$U/$^{232}$Th chains. Facilities with a similar background have reached sensitivities of $\mathcal{O}(50)$\,\textmu Bq/kg for sample masses $>$100\,kg and measurement times $>$50\,days~\cite{heusser06,aprile11}. In a $\sim$51\,g meteorite sample measured for $\sim$18\,days we were able to detect short-lived cosmogenic isotopes like $^{54}$Mn and $^{22}$Na, 4.6\,y after the meteorite fall. GeMSE will therefore help to address important questions in meteoritics like the average fall rate. Furthermore, it will be used for the selection of radiopure materials and components for rare-event searches in astroparticle physics, such as the next generation dark matter experiments XENONnT \cite{aprile14,aprile16} and DARWIN \cite{darwin01,schumann15}.


\acknowledgments

This project is supported by the interdisciplinary SNF grant number 152941, by the Albert Einstein Center for Fundamental Physics and the University of Bern. 
For the consumed computing resources on SWITCHengines we acknowledge the support from swissuniversities.
We thank the mechanical workshop of LHEP Bern for their support and the group of Laura Baudis at the University of Zurich for useful discussion.
We are grateful to S\"onke Szidat from the Chemistry Department of the University of Bern and Andrea Lazzaro from the Technical University of Munich for the screening of the Pb samples. We also thank Thomas Marti and Philipp Steinmann from the Bundesamt f\"ur Gesundheit for providing the CBSS2 source. 


\end{document}